\documentclass{emulateapj}
\pdfoutput=1 

\newcommand{\bLp}{\ensuremath{{\mathit{L}'}}}

\newcommand{\bKs}{\ensuremath{{\mathit{K_s}}}}
\newcommand{\betaPic}{$\beta$\,Pic}
\newcommand{\betaPicb}{$\beta$\,Pic\,b}

\begin{document}

\title{Orbital Constraints on the \betaPic\ Inner Planet Candidate with Keck Adaptive Optics}
\shorttitle{Constraining \betaPicb\ with Keck AO}

\author{Michael P. Fitzgerald\altaffilmark{1,2}}
\altaffiltext{1}{Michelson Fellow; Institute of Geophysics and Planetary Physics, Lawrence Livermore National Laboratory, L-413, 7000 East Ave., Livermore, CA 94550}
\altaffiltext{2}{Department of Physics and Astronomy, UCLA, Los Angeles, CA 90095-1547}
\email{mpfitz@ucla.edu}

\author{Paul G. Kalas\altaffilmark{3} and James R. Graham\altaffilmark{3}}
\altaffiltext{3}{Astronomy Department, University of California, Berkeley, CA 94720, USA}

\shortauthors{Fitzgerald et al.}

\begin{abstract}
A point source observed 8\,AU in projection from $\beta$\,Pictoris in \bLp\ (3.8\,\micron) imaging in 2003 has been recently presented as a planet candidate.  Here we show the results of \bLp-band adaptive optics imaging obtained at Keck Observatory in 2008.
We do not detect \betaPicb\ beyond a limiting radius of 0.29\arcsec, or 5.5\,AU in projection, from the star. 
If \betaPicb\ is an orbiting planet, then it has moved $\geq 0.12$\arcsec\ (2.4\,AU in projection) closer to the star in the five years separating the two epochs of observation.
We examine the range of orbital parameters consistent with the observations, including likely bounds from the locations of previously inferred planetesimal belts.  We find a family of low-eccentricity orbits with semimajor axes $\sim 8$--9\,AU that are completely allowed, as well as a broad region of orbits with $e \lesssim 0.2$, $a \gtrsim 10$\,AU that are allowed if the apparent motion of the planet was towards the star in 2003.  We compare this allowed space with predictions of the planetary orbital elements from the literature.
Additionally, we show how similar observations in the next several years can further constrain the space of allowed orbits.  Non-detections of the source through 2013 will exclude the interpretation of the candidate as a planet orbiting between the 6.4 and 16\,AU planetesimal belts.
\end{abstract}


\slugcomment{ApJL accepted}
\maketitle

\section{INTRODUCTION}\label{sec:intro}

A variety of observations support the hypothesis that A5V star \betaPic\ hosts an orbiting planet.
\citet{smith&terrile84} noted that the dust distribution in the scattered-light disk should be centrally cleared or else significant extinction would be observed due to the edge-on disk orientation.  A clearing within 100\,AU radius was inferred by observations of scattered light and thermal emission that probed closer to the star~\citep{golimowski_etal93, lagage&pantin94, kalas&jewitt95}.  Through simulations,~\citet{roques_etal94} demonstrated that central clearings within debris disks are possible due to mean motion resonances of planets with $M_p>5~M_\earth$.
Spectroscopic observations revealed transient, redshifted absorption features, which are interpreted as rapidly evaporating comets initially located at a 4:1 resonance with a hypothetical planet~\citep{beust&morbidelli00}.
A significant brightening and dimming of the photometric light curve in 1981 is modeled as a planet transit event which could be consistent with an 8\,AU orbit if the planetary radius  $R_p>0.2~R_\star$~\citep[]{lecavelier_etal97}.
Finally, an apparent warp in the inner disk midplane has been attributed to a planetary perturbation, implying that the orbit is not coplanar with the extended dust disk~\citep{burrows_etal95, mouillet_etal97, heap_etal00, augereau_etal01, golimowski_etal06}.

The most recent evidence for a planet orbiting \betaPic\ comes from~\citet{lagrange_etal09}, who present \bLp\ images from 2003 November showing a point source at projected stellar separation $r=0.411\arcsec\pm0.008\arcsec$ at $\mathrm{PA}=31.8\degr\pm1.3\degr$.  They find the apparent brightness of the source, $m_\bLp=11.2\pm0.3$\,mag, is consistent with a 6--12\,$M_J$ planet at the stellar age~\citep[$12^{+8}_{-4}$\,Myr;][]{zuckerman_etal01}.  Assuming a heliocentric distance of 19.3\,pc, the projected separation of \betaPicb\ corresponds to $7.93\pm0.15$\,AU.  Considering this as the minimum semimajor axis of a planet circularly orbiting an $M_\star=1.75$\,$M_\sun$ star~\citep{crifo_etal97}, then the period $P\geq16.9$\,yr.

\section{OBSERVATIONS \& DATA REDUCTION}\label{sec:obs}

\subsection{Observations and Basic Calibration}\label{subsec:obs}

We observed \betaPic\ with the ``narrow'' ($\approx$10\,mas\,pix$^{-1}$) mode of the NIRC2 camera on the Keck II telescope on 2008 December 2.
Exposures consisted of 100 coadds of 53\,ms integrations in correlated double sampling of a 512$\times$512 pixel subarray.  We integrated 667.8\,s on-target in the \bLp\ band ($\lambda_0=3.8$\,\micron, $\Delta\lambda=0.7$\,\micron).
The adaptive optics (AO) loop was closed with \betaPic\ serving as its own wavefront reference.
Despite the low elevation and partial vignetting by the lower edge of the dome shutter, we achieved moderate AO correction, with a diffraction-limited core and first Airy ring visible in individual exposures.
The innermost portion of the PSF core saturated the detector.
The average airmass (AM) was 3.1.
Telluric and instrumental thermal backgrounds are significant in \bLp\ band.  We alternated between \betaPic\ and dithered sky exposures (with equivalent readout configuration) every 25 integrations for a total blank-sky integration of 662.5\,s.
The field rotator was placed in ``vertical angle'' mode, which keeps the point-spread function (PSF) orientation fixed on the detector while the field rotates at the parallactic rate.
The \betaPic\ exposure sequence spanned 26.1$\degr$ of parallactic rotation.

We observed the star FS\,13~\citep[$m_\bLp=10.10\pm0.03$;][]{leggett_etal03} to facilitate photometric calibration.
The low elevation of the \betaPic\ observations relative to those of FS\,13 (mean $\mathrm{AM}=1.1$) affects our calibration strategy.  We must account for the increase in atmospheric extinction as well as the degraded seeing and AO correction of the \betaPic\ measurements.  The vignetting during those exposures further reduced the throughput.
Our goal is faint point source detection, so we focus on calibration in apertures having sizes similar to the PSF core.  We measure $\mathrm{FWHM}=88.5$\,mas in the reduced FS\,13 image, and adopt a nominal circular photometric aperture of this diameter for photometric calibration and subsequent point-source sensitivity analysis.
We derived the photometric zero point for this aperture, accounting for extinction, degraded Strehl ratio, and vignetting, by scaling the image of FS\,13 to match \betaPic.  We extracted a 1\arcsec$\times$1\arcsec\ box around each star and masked the saturated portion of \betaPic's PSF core.  We least-squares fit using the difference between the scaled image of FS\,13 and \betaPic.
Given $m_\bLp=3.47$\,mag for \betaPic~\citep{koornneef83}, we derive a photometric zero point of 21.44\,mag (defined at 1\,DN\,s$^{-1}$) for these apertures.  This gives a scaling to monochromatic flux density of $6.610\times10^{-7}$\,Jy\,s\,DN$^{-1}$~\citep{tokunaga&vacca05}.

We calibrated the detector orientation with 1.6\,\micron\ narrow-band observations of the $\sim6$\arcsec-separation binary HD\,56986.  We found a 0.42\degr\ offset between the observed orientation and the ephemeris computed from the orbital elements in~\citet{hartkopf_etal01}, and used this offset to correct the orientation of the \betaPic\ exposures.
For scale, we adopt 0.009963\,arcsec\,pix$^{-1}$ as measured by~\citet{ghez_etal08}; the value that we derived from the binary was 1\% smaller.

\subsection{PSF Subtraction and Point-Source Detection Sensitivity}\label{subsec:psfsub}

We subtracted the stellar PSF using the angular differential imaging algorithm of~\citet{lafreniere_etal07}.
For each image in the exposure sequence, one constructs a stellar PSF reference using an ensemble of other images in the sequence.  The reference image is formed in patches, with the values in each patch determined by least-squares fitting a linear combination of images in the ensemble.  In each patch, the selection of ensemble images is restricted to those in which the field-of-view has sufficiently rotated to prevent self-subtraction of point-source companions while maintaining fidelity in stellar-speckle suppression.  After constructing a reference PSF for each image, the subtraction residuals are rotated to a common sky orientation and combined.

We processed the images prior to PSF estimation.  For stellar centroid estimation, we used a method similar to that outlined by~\citet{marois_etal06}.  After choosing one image as a reference and high-pass filtering it, we obtained relative centroids by cross-correlating it with the other images.  The reference image centroid was computed via cross-correlation with a version of itself rotated by 180\degr.  After centroiding, we subtracted a radial profile from every image, using a robust mean in each annulus to avoid biases from diffraction spikes.  Finally, we applied a high-pass filter using a boxcar median, 30\,pix on a side.

Evaluation of PSF-subtraction performance requires one to both estimate the residual noise level and gauge the suppression of signals from point sources.
We assessed residual noise by examining the distribution of fluxes from 88.5-mas-diameter circular apertures placed in annuli around the star.
Since speckles in AO-corrected images do not follow Gaussian statistics~\citep[e.g.][]{fitzgerald&graham06}, the distribution of residual flux at a given location may not be Gaussian even after combination of several images~\citep{marois_etal08}.  With this caveat in mind, our confidence limits are based on the sample standard deviation of aperture photometry in each annulus, presuming Gaussian statistics.

To assess the suppression of signal of faint sources in the data, we inserted artificial point sources in the images prior to processing.
These sources' PSFs are determined by the normalized image of FS\,13.
We then applied the same boxcar filter as in the pre-processing step above, and
scaled the sources to 11.2\,mag (the candidate flux).
In the outer regions of the image, sources are placed in a radial spoke pattern, with spokes separated by 60\degr\ and sources placed every 20 pixels.  To prevent biases from crowding close to the star, sources at radius $r$ are required to maintain an arcwise separation $r\Delta\theta\geq50$\,pix.
We then measured the flux recovered by the algorithm at each radius with aperture photometry.  Sources closer to the star undergo greater suppression.  We least-squares fit an \textit{ad hoc} function to the fraction of recovered flux as a function of radius,
\begin{equation}
f(r) = a\left\{1+b\left[\exp(r/c)-1\right]^{-1}\right\}^{-1}.
\end{equation}
This function provides a reasonable fit to the recovered artificial source fluxes from 0.2\arcsec--2.5\arcsec.  Here, $a$ is the fraction of flux passed at large radii $r$, $c$ is a transition radius where throughput drops, and $b$ governs the sharpness of the curve.  To illustrate, at $r=(0.3\arcsec,0.6\arcsec,1.0\arcsec,1.5\arcsec,2.0\arcsec,2.5\arcsec)$, we find $f(r)=(0.42,0.63,0.77,0.86,0.91,0.93)$, respectively.
We then computed the point-source sensitivity as a function of radius after dividing the noise profile by the signal suppression function $f(r)$.

The subtraction algorithm has several parameters, which we tuned through trial and error to maximize the point-source sensitivity close to the star.  \citet{lafreniere_etal07} discuss these parameters in detail.
We adopted $N_d=88.5$\,mas for the resolution element width, $N_d=0.5$ for the minimum rotation for an exposure to be usable as a reference, and $N_A=100$ elements in the optimization area.
The subtraction zone widths were $dr=1.5\times W$ for 0--125\,pix, and $5\times W$ beyond.
We adopted $g=1$ for the optimization region aspect ratio.

\section{RESULTS \& ANALYSIS}\label{sec:analysis}

\begin{figure*}
\plotone{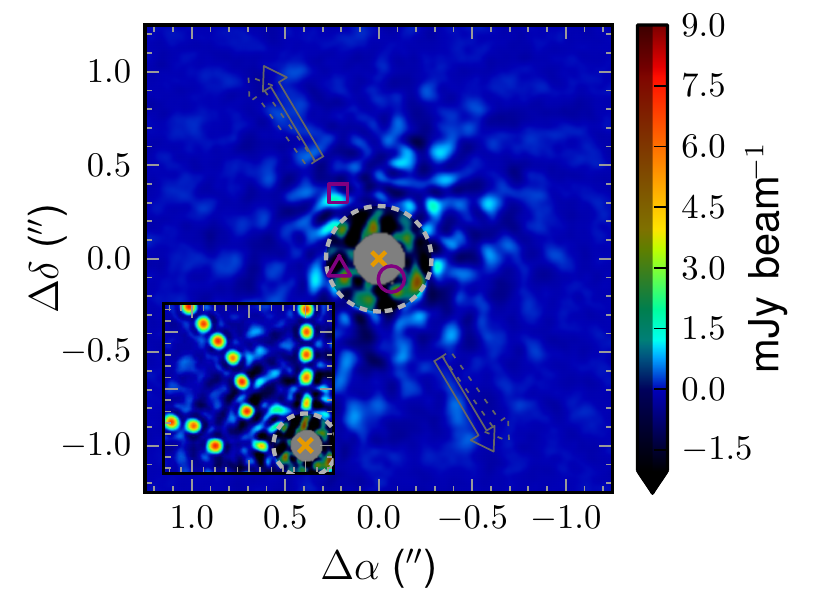}
\caption{Keck \bLp-band PSF-subtracted image of \betaPic.  North is up, east is left.  The dotted circle indicates the extent of our 3-$\sigma$ sensitivity limit ($m_\bLp = 11.5$\,mag; cf. Fig.~\ref{fig:sensitivity}), and the region interior to this limit is shaded.  No point sources are detected outside this region.  A square northeast of the star marks the position of the planet candidate detected by~\citet{lagrange_etal09} at $r \simeq 0.41\arcsec$, PA$\simeq 32\degr$ in 2003 Nov.  A triangle to the lower left of the star indicates the expected position of the source in our 2008 Dec data if it were a background object, though this hypothesis cannot be ruled out since such a source would not be detectable in this region.  The circle to the lower right is the expected position of the companion if it were on a circular orbit and located at maximum elongation in 2003 Nov, which again is not ruled out by the current data.  Solid arrows indicate the orientation of the outer disk midplane~\citep{kalas&jewitt95}, while dotted ones mark the position angle of the secondary disk inferred by~\citet{golimowski_etal06}.  The inset shows output of the PSF-subtraction procedure after the insertion of artificial 11.2\,mag sources (\S\ref{subsec:psfsub}).}\label{fig:image}
\end{figure*}

Our final PSF-subtracted image is shown in Figure~\ref{fig:image}.
The apparently point-like $m_\bLp=11.2\pm0.3$\,mag object detected by~\citet{lagrange_etal09} has not yet been confirmed in the literature.  In this section, we consider the implications of our \bLp\ image for the presence of a bound companion.  Before our analysis of companion orbits, we briefly consider the alternatives of a persistent feature in the circumstellar disk and an unbound background object.

\subsection{Expected Source Locations}\label{subsec:srclocs}
At the 2003 source position~\citep[$r=411\pm8$\,mas, $\mathrm{P.A.}=31.8\degr\pm1.3\degr$;][]{lagrange_etal09}, we do not detect any point sources consistent with the flux measured in the discovery image.
Figure~\ref{fig:image} shows an elongated peak ($r=395$\,mas, $\mathrm{P.A.}=38\degr$) located near this discovery position.  We measure $m_\bLp=12.3$\,mag in an 88.5\,mas aperture, which is 3.1$\times$ above the residual 1-$\sigma$ sample standard deviation at that radius~(\S\ref{subsec:psfsub}) and 1.1\,mag fainter than the putative planet detected in 2003.  We do not have sufficient confidence to assess this as a signal detection given the likely deviation from Gaussian statistics noted above.
If the 2003 source was due to a localized density enhancement in the circumstellar disk, then it has not persisted to be detectable at the same peak brightness along the same line of sight in the 2008 data.

\begin{figure}
\centering\includegraphics{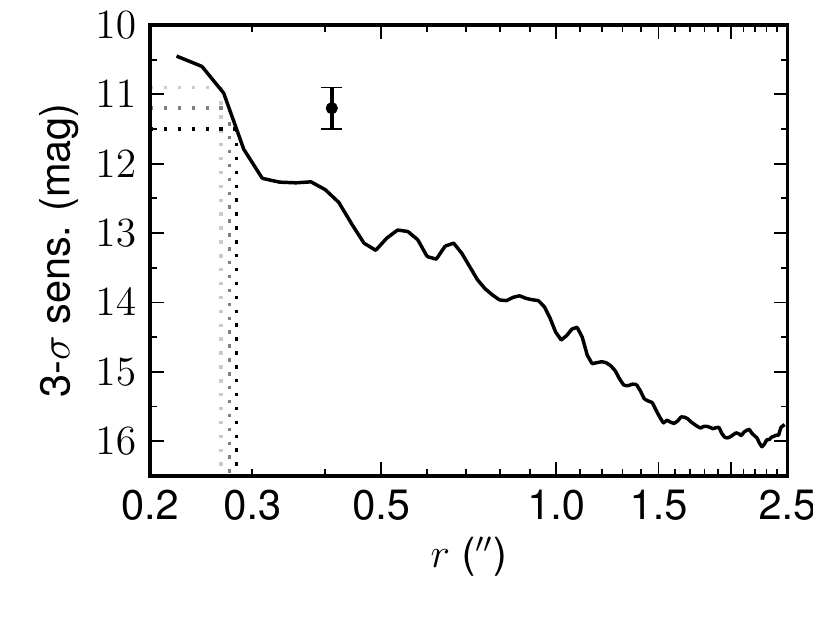}
\caption{Keck 3-$\sigma$ \bLp-band sensitivity limit for \betaPic.  This curve was derived from the standard deviation of 88.5-mas-diameter aperture photometry, corrected for the radial dependence of the throughput of the roll-subtraction algorithm~(\S\ref{subsec:psfsub}).  We show the detection of an $11.2 \pm 0.3$\,mag source at $r \simeq 0.41$\arcsec\ by~\citet{lagrange_etal09}.  For this range of source brightness, we show the corresponding projected separations, and adopt  that of the darkest vertical dotted line ($r\geq 0.29$\arcsec) as an upper limit to the allowed projected separation of the source.}\label{fig:sensitivity}
\end{figure}

We do not detect any point source at the position expected if the~\citet{lagrange_etal09} detection was due to an unlikely background object; however this position falls within our sensitivity limit and is therefore not significant.
The sensitivity curve shown in Fig.~\ref{fig:sensitivity} shows that we would likely detect an 11.2\,mag object beyond 0.29\arcsec.  This limit is also shown as a dotted circle in Fig.~\ref{fig:image}.
A background source would be detectable at this sensitivity beginning in 2010 Jan.

Our image shows no 11.2-mag point sources outside the 0.29\arcsec\ limit.
In Fig.~\ref{fig:image}, we mark the expected location if the companion was observed at maximum elongation in an edge-on circular orbit in 2003.  This location is within our sensitivity limit, so we are unable to rule out a body in such an orbit.

\subsection{Orbital Constraints}\label{subsec:orbits}
Our non-detection constrains other possible orbital configurations for the \betaPic\ planet candidate.
Given the detection at the best-estimate position of~\citet{lagrange_etal09} in 2003, what orbital parameters are allowed with our exclusion of sources at $r>0.29$\arcsec\ in the 2008 epoch?
For simplicity, we restrict our consideration to edge-on geometries ($i=90\degr$).
Atomic gas emission, resolved in both space and velocity, shows quasi-Keplerian rotation with a redshifted NE ansa.  We presume the candidate planet shares this orbital direction and define the longitude of the ascending node $\Omega$ to correspond to the P.A. of the 2003 detection.
The remaining free elemental parameters are semimajor axis $a$, eccentricity $e$, and argument of periastron $\omega$.  If we define the direction of $\Omega$ to be the $x$ axis, we then have the $y$ axis pointing away from the star along the line of sight and $\omega$ being measured from $x$ through $y$ (i.e. $\arg x+iy$).
We constructed a grid in $(a,e,\omega)$ and computed the possible projected positions ($x$ coordinates) of the source in the latter epoch.  In general, there are two solutions for the anomaly of the 2003 epoch at each $(a,e,\omega)$ owing to the near and far points of intersection between an edge-on orbit and the line of sight.  These solutions correspond to two possible positions in 2008.

\begin{figure}
\centering\includegraphics{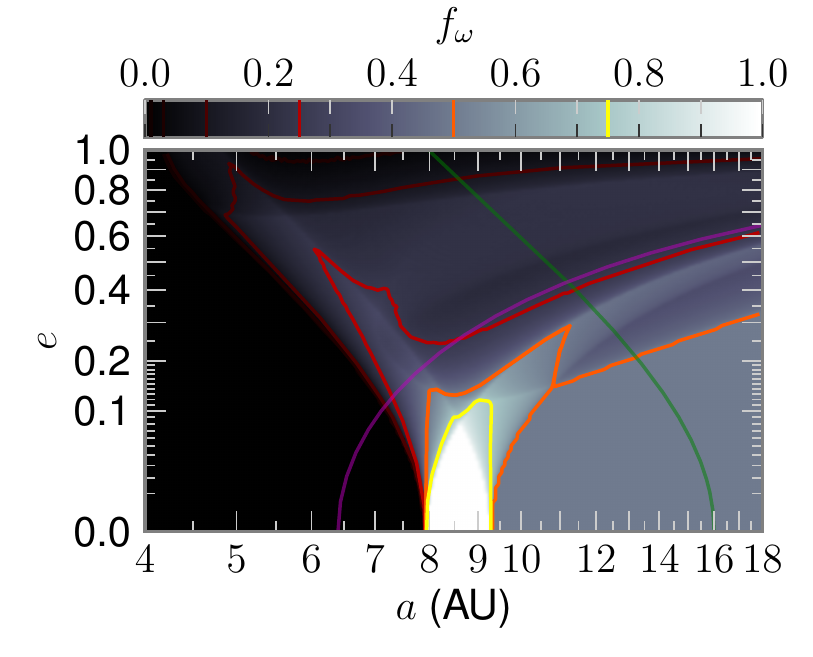}
\caption{Map of orbital parameters allowed by the \bLp\ observations assuming the candidate is a bound companion.  At each position in $(a,e)$, we show the fraction of orbital periastron longitudes $f_\omega$ that are allowed given a detection in 2003 and our 2008 non-detection.  The magenta curve shows the constraint $r_\mathrm{peri} = 6.4$\,AU, while the green shows $r_\mathrm{ap} = 16$\,AU.  The region of intersection below both of these curves is the range of orbits that are not allowed to cross the corresponding~\citet{okamoto_etal04} belt locations.}\label{fig:orbits}
\end{figure}

The range of orbital parameters allowed by the data is shown in Figure~\ref{fig:orbits}.  The shading corresponds to the fraction of allowed orbits, $f_\omega$ over the full range of $\omega$.
In the absence of prior information on $\omega$, this fraction corresponds to probability of an allowed orbit given $a$ and $e$.
A key feature of the allowed parameter space
is a family of low-eccentricity orbits with semimajor axes $\sim8$--9\,AU that are completely allowed by the observations (i.e. no constraints on $\omega$).
The completely disallowed region on the left edge of the graph (where $f_\omega=0$) corresponds to the requirement that the apastron distance must be at least as great as the separation observed in 2003.
On the other side of the Figure, there is a broad region of parameter space ($e\lesssim0.2$, $a\gtrsim10$\,AU) corresponding to wider orbits where the planet has a larger physical separation but smaller apparent separation in the current data.  Here, $f_\omega=50\%$, as half of the orbits (the near solutions) have the apparent separation increasing with time.

We further constrain the allowable orbits by considering the locations of planetesimal belts inferred from peaks in the crystalline silicate fraction observed with spatially resolved mid-infrared spectroscopy~\citep{okamoto_etal04}.
By requiring non-crossing orbits, a belt at 6.4\,AU constrains the periastron distance,
while the inferred belt at 16\,AU constrains apastron.  As noted by~\citet{freistetter_etal07}, these belt locations may be uncertain by a few AU due to relatively coarse spatial sampling.  We show these peri- and apastron constraints in Fig.~\ref{fig:orbits}.  Together, they define a wedge-shaped region in $(a,e)$ space and require $e\lesssim0.4$.

\begin{figure}[ht!]
\centering\includegraphics{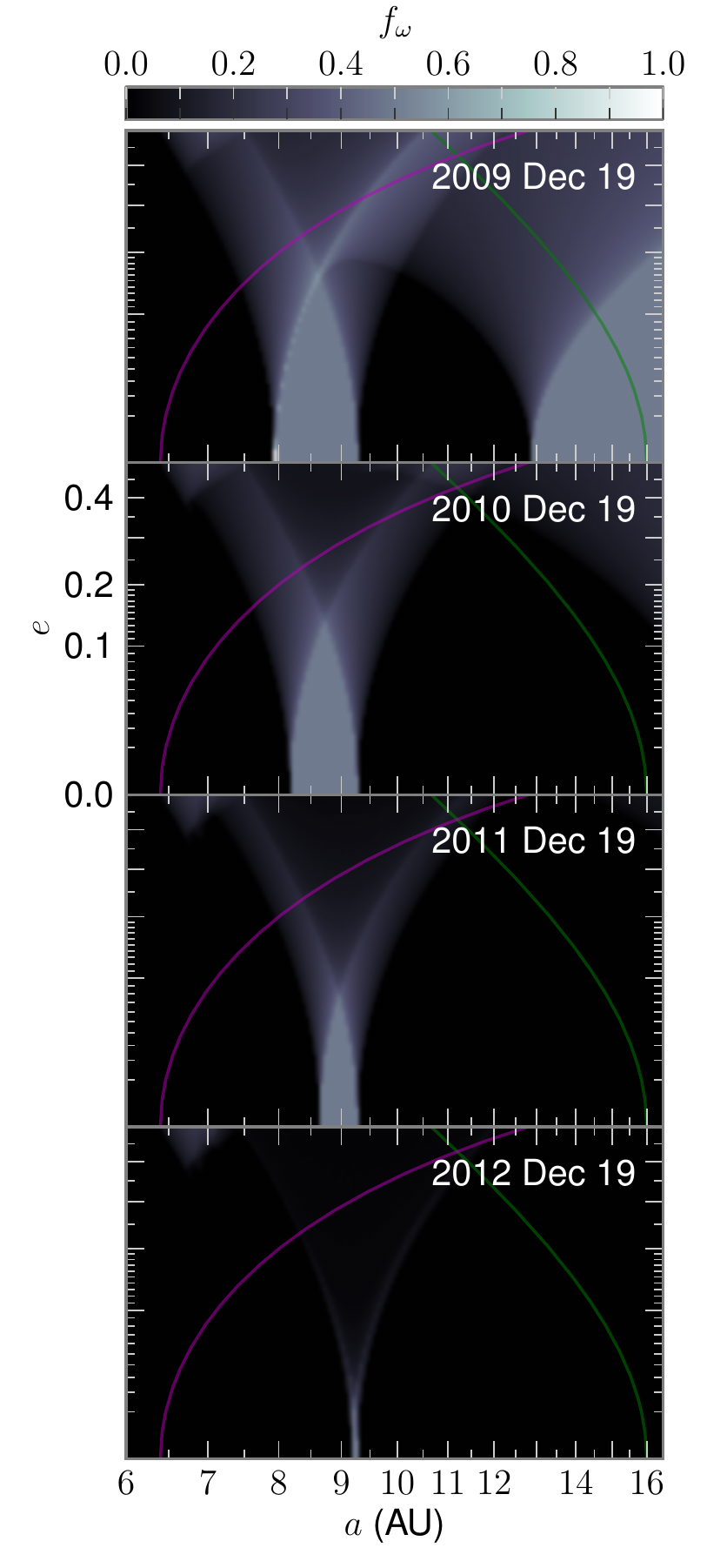}
\caption{Map of orbital parameters for the candidate allowed by hypothetical future non-detections.  As in Fig.~\ref{fig:orbits}, these panels depict the $f_\omega$ at each $(a,e)$; however in this case we presume non-detection at the same sensitivity each year through the date shown in each panel.  The candidate detection of~\citet{lagrange_etal09} can only be ruled out of this parameter range if non-detections continue through 2013.}\label{fig:futureorbits}
\end{figure}

\citet{boccaletti_etal09} present non-detections $>0.5$\arcsec\ in the \textit{H} band and $>0.4$\arcsec\ in $K_s$ in the 2004 November.  Due to the modeling uncertainties in translating these non-detections to \bLp-band equivalents, we do not incorporate these results into our orbital phase space constraints.

Previous works have made predictions for the orbits of putative planets, which we compare to our presently allowed space.
Orbital predictions have been derived from the velocity distribution of transient gas absorption events hypothesized to arise from star-grazing comet-like bodies disrupted from their parent belt by mean-motion resonance interactions with a perturbing planet~\citep{beust_etal98, beust&morbidelli00}.  For a planet in the 4:1 resonance, these authors predict $e\gtrsim0.6$, $5\,\mathrm{AU}\lesssim{a}\lesssim20\,\mathrm{AU}$, and a periastron longitude, measured from the line of sight, between -80\degr\ and -60\degr.  In the coordinate system defined here, this translates to $-170\degr\lesssim\omega\lesssim-150\degr$.
Within these constraints, the current data do not place strong limitations on the range of allowable orbits.  In this region, we find that orbits with apastra $r_\mathrm{ap}\gtrsim9.4$\,AU must correspond to the far solutions, i.e. those with a true anomaly between 0\degr\ and 90\degr\ in the 2003 observation.

\citet{mouillet_etal97} have studied how a planetary perturber can influence the apparent warp in the disk midplane.  We note that the nature of this structure is unclear, as it may be due to an inclined secondary disk~(\citealt{golimowski_etal06}; however see also~\citealt{boccaletti_etal09}).  As~\citet{lagrange_etal09} note, a mass range of 6--13\,$M_\mathrm{J}$ estimated with atmosphere models, coupled with the age-mass-semimajor-axis constraint of~\citet{mouillet_etal97}, corresponds to $a\sim7.6$--9.7\,AU.  Our data in Fig.~\ref{fig:orbits} are largely consistent with this range.
\citet{freistetter_etal07} have numerically studied the possible architectures of the planetary system given the belt structure inferred by~\citet{okamoto_etal04} as well as the predictions for planetary perturbers that can account for falling evaporating bodies and the warp.  The favored single-planet scenario has a 2\,$M_\mathrm{J}$ planet at $a=12$\,AU with $e\lesssim0.1$.  The orbital parameters are consistent with the current data, though the mass of such a planet must be reconciled with the apparent brightness and age.

Most recently,~\citet{lagrange_etal09b} present non-detections of the candidate planet in \bLp\ and \bKs\ data from 2009 Jan and Feb, respectively.  Restricting the comparison to \bLp, we achieve similar sensitivity levels.  The data were obtained at similar epochs and there is a general concordance between the resulting conclusions, though we explicitly consider non-circular orbits. 
Both~\citet{lecavelier&vidal-madjar09} and~\citet{lagrange_etal09b} consider whether the candidate is consistent with causing the transit-like event observed in 1981.  Again restricting consideration to the \bLp\ observations, both groups find consistency with orbits having semimajor axes of $\sim8$ or $\sim17$\,AU.  As can be seen in Fig.~\ref{fig:orbits}, our non-detection is consistent with both families of orbits.

Additional observations of \betaPic\ in the mid-term can further constrain the nature of the candidate.  In Fig.~\ref{fig:futureorbits} we examine the fraction of allowed orbits at each $(a,e)$ in between the 6.4 and 16\,AU planetesimal belts presuming no planet is detected outside of 0.29\,\arcsec\ in the next several years.
For a non-detection in late 2009, the range of allowed orbits will be significantly constrained (top panel), with the elimination of longer period, near-circular orbits
(with $r_\mathrm{ap}\gtrsim13$\,AU)
occurring the following year.
The 2003 detection can only be judged spurious if, at these sensitivity levels, non-detections continue through 2013.

\section{CONCLUSIONS}\label{sec:conclusions}

We observed \betaPic\ in the \bLp\ band on 2008 Dec 2 and did not detect any point sources.  In our data, the planet candidate observed in 2003 would have been detected if it were as close as 0.29\arcsec\ from the star.
Our sensitivity limit alone does not exclude the possibility that the source is an unlikely background object, nor can it rule out a companion on a circular orbit if it was observed at maximum elongation in 2003.
By mapping the phase space of allowable orbital parameters, and considering the likely location between two previously detected planetesimal belts, we determine that the planet candidate is spurious only if it is not detected in similar \bLp\ observations conducted through 2013.

\acknowledgements
We would like to thank Bruce Macintosh and Herv{\'e} Beust for helpful discussions.
M.P.F. acknowledges support from the Michelson Fellowship Program, under contract with JPL, funded by NASA.  Work at LLNL was performed under the auspices of DOE under contract DE-AC52-07NA27344.
P.G.K. and J.R.G. are supported in part by the NSF Center for Adaptive Optics, managed by the University of California at Santa Cruz under cooperative agreement No. AST-9876783.

{\it Facilities:} \facility{Keck:II (NIRC2)}

\bibliography{}

\end{document}